# Layer-Resolved Magnetic Proximity Effect in van der Waals Heterostructures

**Authors**
Ding Zhong[1], Kyle L. Seyler[1], Xiayu Linpeng[1], Nathan P. Wilson[1], Takashi Taniguchi[2], Kenji Watanabe[2], Michael A. McGuire[3], Kai-Mei C. Fu[1,5], Di Xiao[4], Wang Yao[6], Xiaodong Xu[1,7*]

**Affiliations**
[1]Department of Physics, University of Washington, Seattle, Washington 98195, USA.
[2]National Institute for Materials Science, 1-1 Namiki, Tsukuba 305-0044, Japan.
[3]Materials Science and Technology Division, Oak Ridge National Laboratory, Oak Ridge, Tennessee, 37831, USA.
[4]Department of Physics, Carnegie Mellon University, Pittsburg, Pennsylvania 15213, USA.
[5]Department of Electrical and Computer Engineering, University of Washington, Seattle, Washington 98195, USA.
[6]Department of Physics and Center of Theoretical and Computational Physics, University of Hong Kong, Hong Kong, China.
[7]Department of Materials Science and Engineering, University of Washington, Seattle, Washington 98195, USA.

*Correspondence to xuxd@uw.edu

**Abstract:** Magnetic proximity effects are crucial ingredients for engineering spintronic[1], superconducting[2], and topological phenomena[3,4] in heterostructures. Such effects are highly sensitive to the interfacial electronic properties, such as electron wave function overlap and band alignment. The recent emergence of van der Waals (vdW) magnets enables the possibility of tuning proximity effects via designing heterostructures with atomically clean interfaces[5-20]. In particular, atomically thin $CrI_3$ exhibits layered antiferromagnetism, where adjacent ferromagnetic monolayers are antiferromagnetically coupled[5]. Exploiting this magnetic structure, we uncovered a layer-resolved magnetic proximity effect in heterostructures formed by monolayer $WSe_2$ and bi/trilayer $CrI_3$. By controlling the individual layer magnetization in $CrI_3$ with a magnetic field, we found that the spin-dependent charge transfer between $WSe_2$ and $CrI_3$ is dominated by the interfacial $CrI_3$ layer, while the proximity exchange field is highly sensitive to the layered magnetic structure as a whole. These properties enabled us to use monolayer $WSe_2$ as a spatially sensitive magnetic sensor to map out layered antiferromagnetic domain structures at zero magnetic field as well as antiferromagnetic/ferromagnetic domains near the spin-flip transition in bilayer $CrI_3$. Our work reveals a new way to control proximity effects and probe interfacial magnetic order via vdW engineering[21].

**Main Text:**

At the interface formed by a magnetic and nonmagnetic material, the magnetic order can drastically influence the properties of the nonmagnetic component[22,23], which can expose new functionalities absent from the individual materials. This proximity effect is usually short-ranged due to the finite extension of the electronic wavefunctions across the interface. Thus, vdW materials, which feature atomic thicknesses and form atomically sharp interfaces, are an attractive platform to realize and harness the proximity effect.

Indeed, emergent phenomena have been observed in heterostructures formed by monolayer WSe$_2$ and magnetic insulator CrI$_3$ (Fig. 1a)[6,24]. Monolayer WSe$_2$ is a non-magnetic semiconductor with coupled spin-valley physics[25]. By interfacing it with CrI$_3$, the proximity-induced exchange field gives rise to spontaneous valley excitonic Zeeman splitting. In addition, a type-II band structure forms at the WSe$_2$/CrI$_3$ heterostructure interface with the lowest conduction band in CrI$_3$. Since the relevant bands in CrI$_3$ are spin-polarized, this type-II band structure facilitates spin-dependent charge transfer between WSe$_2$ and CrI$_3$ (Fig. 1b), leading to large spontaneous exciton valley-spin polarization. These findings suggest WSe$_2$/CrI$_3$ as a promising platform to perform magnetic proximity effect studies. However, for 10 nm thick CrI$_3$, as studied in Ref. 6, the magnetic structure is too complicated to unravel several fundamental issues. For instance, the magnetic proximity effect involves both real and virtual electron hopping at the heterostructure interface; it is unclear how the two hopping processes manifest at the heterostructure interface and beyond. It remains elusive how the observed valley dynamics in WSe$_2$ connect to the magnetic states in CrI$_3$. Such knowledge is important for understanding the magnetic properties of CrI$_3$ (e.g., magnetic domains) and engineering magnetic vdW heterostructures to harness the proximity effect for new functionalities.

Here we report a magneto-optical spectroscopy study of the proximity effect in heterostructures formed by monolayer WSe$_2$ with either bilayer or trilayer CrI$_3$. We employed polarization-resolved magneto-photoluminescence to measure the effect of CrI$_3$ proximity on the valley-spin dynamics of WSe$_2$. Combined with reflective magnetic circular dichroism (RMCD) experiments, which probe the magnetic order of CrI$_3$, we show that spin-dependent charge transfer between WSe$_2$ and CrI$_3$ is dominated by the interfacial CrI$_3$ layer magnetization, while the proximity exchange field has a strong dependence on the layered magnetic structure as a whole. Building on this layer-resolved effect, we used monolayer WSe$_2$ as a magnetic sensor to probe the magnetic domains in bilayer CrI$_3$, which is a challenge for conventional techniques, such as RMCD, due to the vanishing magnetization of the AFM order. At zero magnetic field, we uncovered both reconfigurable and pinned layered AFM domain walls. Furthermore, near the magnetization-flip transitions, we observed the evolution of AFM/FM domain walls as a function of the magnetic field.

The WSe$_2$/CrI$_3$ heterostructures were fabricated by mechanical transfer of individual exfoliated WSe$_2$ and CrI$_3$ flakes in a glovebox (Methods). A device schematic is shown in Fig. 1a, where WSe$_2$ is on top of CrI$_3$. The excitation laser was fixed at 1.96 eV, and the optical axis and applied magnetic field are perpendicular to the sample plane. For measurement of monolayer WSe$_2$ photoluminescence, we performed co-circular polarized excitation and detection (either $\sigma^+/\sigma^+$ or $\sigma^-/\sigma^-$) to read out the valley exciton information.

We first present the results from a monolayer WSe$_2$/trilayer CrI$_3$ heterostructure. Fig. 1c shows circular polarization-resolved photoluminescence spectra at 15 K and zero magnetic field, which is dominated by positively charged trion emission[6], consistent with type-II band alignment. The $\sigma^+/\sigma^+$ photoluminescence (red curve) is stronger than $\sigma^-/\sigma^-$ (blue curve); this spontaneous circularly polarized photoluminescence demonstrates the breaking of valley degeneracy and thus time reversal symmetry of monolayer WSe$_2$ by proximity to the magnetic trilayer CrI$_3$. As discussed in Ref. 6, when the photoexcited electron spin in WSe$_2$ has the same orientation as the CrI$_3$ magnetization, charge transfer is allowed (Fig. 1b). Otherwise, it is suppressed. This spin-dependent charge transfer from WSe$_2$ to CrI$_3$ gives rise to strong circularly polarized photoluminescence (Fig. 1c).

To investigate the relationship between the proximity effect and magnetic states, we measured polarization-resolved photoluminescence and RMCD as a function of applied magnetic field at 15 K. In labeling the magnetic states of CrI$_3$ below, we count the CrI$_3$ layers from top to bottom, where the top layer interfaces with WSe$_2$. We quantify the photoluminescence polarization as $\rho = (I_{\sigma^+} - I_{\sigma^-})/(I_{\sigma^+} + I_{\sigma^-})$, where $I_{\sigma^+}$ ($I_{\sigma^-}$) represents photoluminescence intensity with co-$\sigma^+$ (co-$\sigma^-$) excitation and detection. In addition to spin-dependent charge transfer, the magnetic exchange field introduces an excitonic valley Zeeman splitting, which is defined as $\Delta = E_{\sigma^+} - E_{\sigma^-}$. Here, $E_{\sigma^+}$ ($E_{\sigma^-}$) is the peak energy of $\sigma^+/\sigma^+$ ($\sigma^-/\sigma^-$) photoluminescence from K (-K) valley trions (Supplementary Information S1). Figs 1d-f plot $\rho$, $\Delta$, and RMCD signal as a function of magnetic field, respectively. Consistent with previous reports, the RMCD signal in Fig. 1f shows three transitions in a given field sweep (at about $\pm 1.6$T and $\pm 0.2$T), with each corresponding to a flip in layer magnetization. The magnetic states, which consist of two fully spin-polarized states ↑↑↑ and ↓↓↓, and two layered AFM states ↑↓↑ and ↓↑↓, are indicated in Fig. 1f.

As shown in Figs 1d, e, the evolution of magnetic states as a function of magnetic field is also manifested in both $\rho$-H and $\Delta$-H traces. We first focus on the $\rho$-H trace and sweep the magnetic field up from large negative field where the CrI$_3$ trilayer is fully polarized (↓↓↓). Due to Hund's coupling in the Cr ions, photoexcited electrons with spin down in WSe$_2$ transfer more efficiently to CrI$_3$, resulting in stronger $\sigma^+$ polarized PL. As the magnetic field increases beyond -1.6 T, CrI$_3$ transitions from ↓↓↓ into ↓↑↓, with the middle layer flipping its magnetization. Correspondingly, $\rho$ decreases slightly, by about 16%. Further increasing the magnetic field to be above 0.2 T causes CrI$_3$ to transition into ↑↓↑ with the top layer flipping its magnetization. As a result, $\rho$ sharply changes from positive to negative since electrons with spin up are now favored to transfer from WSe$_2$ to CrI$_3$. When the magnetic field is larger than 1.6 T, CrI$_3$ becomes fully spin polarized (↑↑↑) and $\rho$ reaches maximum negative value. The $\rho$-H trace implies that the spin-dependent charge transfer is dominated by the interfacial top layer, while the middle layer, 0.7 nm away below the interface, has a measurable but much weaker effect.

In contrast to $\rho$, the proximity exchange field, and thus the induced valley Zeeman splitting $\Delta$ has a distinct dependence on the magnetic states. As shown in Fig. 1e, when CrI$_3$ is the AFM states, $\Delta$ is much larger than it is for the fully spin-polarized states. This is quite surprising since the proximity exchange field is from a short-range interaction, and thus, it is expected to be dominated by the magnetization in the top layer. The pronounced difference in the proximity exchange field between the fully spin polarized and AFM states (e.g. ↓↓↓ and ↓↑↓) is unexpected.

This observation is further corroborated by measurements on monolayer WSe$_2$/bilayer CrI$_3$ heterostructures at 1.6K. Fig. 2a shows a schematic of the bilayer heterostructure. The RMCD signal in Fig. 2b is typical of bilayer CrI$_3$, showing two AFM states (↑↓ and ↓↑) and two fully spin-polarized FM states (↑↑ and ↓↓). Fig. 2c shows the photoluminescence intensity plot of co-$\sigma^+$ (left) and co-$\sigma^-$ (right) excitation and detection as a function of magnetic field and photon energy. Sharp changes are identified near the transition of CrI$_3$'s magnetic states. The extracted $\rho$-H curve in Fig. 2d shows that $\rho$ reaches maximum when CrI$_3$ is in the fully spin-polarized states and has a slight decrease in the CrI$_3$ AFM states. In contrast, in Fig. 2e, AFM states produce larger valley Zeeman splitting than the FM states.

The distinct behavior between $\rho$ and $\Delta$ can be explained as follows. The photoluminescence polarization $\rho$ is determined by the real hopping of electrons from WSe$_2$ to CrI$_3$ accompanied by energy relaxation. In contrast, the proximity exchange field is primarily from a second-order virtual hopping process that shifts the WSe$_2$ band edge of certain spin species by $\Delta_{c/v} \sim \frac{t^2}{\Delta E}$, where $t$ is the hopping matrix element between WSe$_2$ and CrI$_3$, and $\Delta E$ is their band offset. In the AFM configuration (e.g., ↑↓), the anti-parallel spin alignment of adjacent CrI$_3$ monolayers suppresses the interlayer hopping of charge carriers, which is otherwise significant in the fully spin aligned configuration (e.g., ↑↑). This magnetic-configuration-dependent interlayer hopping between CrI$_3$ layers leads to significantly different band edge energies of the two magnetic configurations. The measured Zeeman splitting in the photoluminescence is the difference between the conduction and valence band shifts: $\Delta = \Delta_c - \Delta_v$. We find that the observed change in $\Delta$ can be reproduced with reasonable choices of $t$ and band offset $\Delta E$ (see Supplementary Information S2), while the precise determination of these parameters requires further experimental (e.g., angle-resolved photoemission spectroscopy) and computational studies.

The sensitive dependence of the proximity effect on the spin structure of CrI$_3$ shown above allows us to use monolayer WSe$_2$ as a magnetic sensor to probe the domain structures and dynamics in layered antiferromagnetic bilayer CrI$_3$. Since bilayer CrI$_3$ has vanishing magnetization in the AFM configuration, RMCD alone cannot probe the domain effects. However, the photoluminescence polarization, $\rho$, is dominated by the interfacial layer, thus providing an excellent probe of magnetization in the top CrI$_3$ layer. Combined with RMCD, which probes the total magnetization of the bilayer, we can construct the layered AFM domains.

Fig. 3a shows the intensity map of $\rho$ (top) and RMCD signal (bottom) of monolayer WSe$_2$/bilayer CrI$_3$ device BD1. The data was acquired after the sample was zero-field-cooled (ZFC) down to 1.6 K. The map of $\rho$ is composed of positive and negative polarization patches, which corresponds to the up and down magnetization domains of the top layer in the CrI$_3$ bilayer. On the other hand, the RMCD map (bottom panel) has nearly zero intensity across the whole heterostructure, showing that bilayer CrI$_3$ is in an AFM ground state. The comparison of both maps reveals the spontaneous formation of ↑↓ and ↓↑ layered AFM domains.

The spontaneously formed layered AFM domain walls can be reconfigured by applying an external magnetic field. For instance, after initializing the bilayer to the ↓↓ state by a large negative magnetic field, we sweep the magnetic field back to zero. The resulting spatial map of $\rho$ in Fig. 3b shows two dominant layered AFM domains, labeled by I and II. We can also initialize the bilayer in the ↑↑ state and then sweep the magnetic field back to zero. The corresponding spatial map of $\rho$, shown in Fig. 3c, is a time reversal of Fig. 3b. Figures 3d and e show $\rho$-H at two selected spatial points in domains I and II, respectively. When the magnetic field sweeps up, at around $-0.5$ T, the magnetic states switch from FM to AFM, $\rho$ drops slightly in domain I (Fig. 3d), while it changes drastically and reverses sign in domain II (Fig. 3e). These data demonstrate that the top layer in CrI$_3$ bilayer flips magnetization first in domain II, while the bottom layer flips first in domain I (see Fig. S3 for additional data). Comparing Figs. 3b&c with Fig. 3a, we can see that among the spontaneously formed layered AFM domains in Fig. 3a, domain II is pinned with a fixed boundary, while others are movable and reconfigured after magnetization initialization. This pinned domain II is likely due to strain introduced by the heterostructure fabrication.

We found that the domain structures vary between devices, likely caused by the uncontrolled heterostructure fabrication process. Figure 4 shows another example (device BD2) measured at 6.6 K. The ZFC domain structure is mainly dominated by two layered AFM domains (Fig. 4a). After the bilayer $CrI_3$ has been initialized by the magnetic field, the layered AFM domains at zero field vanish, and only a single domain exists (Fig. 4b). However, at finite magnetic fields, we observed a layered AFM/FM domain wall. For instance, as the magnetic field increases from zero, a ↑↑ domain forms first at the bottom left corner of the heterostructure at a magnetic field of 0.25 T, while the rest of the sample is still in the ↓↑ state. Distinct ρ-H traces from these two domains are shown in Fig. S4. Compared to BD1, this device has less complicated domain structures at zero magnetic field, which may indicate a better transfer with a more homogeneous strain distribution.

We have also observed complicated, layered AFM/FM domain patterns in a less homogenous sample compared to device BD2. Figure 5 shows spatial maps of both RMCD and ρ on a third monolayer $WSe_2$/bilayer $CrI_3$ device (BD3) as the magnetic field sweeps through the metamagnetic transition at 1.6 K. The RMCD maps in Fig. 5a show faint spotty patterns, indicating spatial inhomogeneity in the transition from ↓↑ to ↑↑. The evolution of domains across this metamagnetic transition is clearer in the spatial maps of ρ (Fig. 5b), which highlights the domain dynamics near spin-flip transition (see Fig. S5 for additional analysis). Fig. S6 shows similar complicated domain effects in BD1. As suggested, the inhomogeneity is likely from strain introduced in the heterostructure transfer process (see Fig. S7).

In summary, our work provides insights into the magnetic proximity effects of a vdW magnetic heterostructure by utilizing the multiple magnetic configurations accessible in layered antiferromagnetic $CrI_3$. This understanding will be important for developing magnetic vdW devices. In addition, we establish the ability to image layered antiferromagnetic domains within a vdW magnet using the spin-valley properties of a monolayer semiconductor, which would be challenging with conventional magnetometry techniques. Furthermore, our observation of possible strain-induced domain behavior highlights the effect of strain in determining magnetic states in vdW magnets. Thus we suggest that strain engineering is a promising direction for future studies to enhance the versatility and controllability of vdW magnets in spin and valleytronics applications.

**Methods**

**Device Fabrication**
Monolayer $WSe_2$ and hBN flakes were exfoliated in the ambient condition and then transported into the glove box. $CrI_3$ flakes were exfoliated onto 90nm silicon oxide/silicon wafer inside the glovebox with < 0.5 ppm $O_2$ and < 0.5 ppm $H_2O$ environment. We identified bilayer $CrI_3$ by optical contrast, which was further confirmed by RMCD in the experiment. Immediately after finding the bilayer, we assembled the heterostructure by polycarbonate-based dry transfer. After hBN encapsulation, the heterostructures were stable in ambient environment, which allowed us to remove it from the glovebox and load it into an optical cryostat for measurements.

**Optical Measurements**
Optical measurements were performed using a setup detailed in Ref. 6. We used a HeNe laser (1.96eV) for both polarization-resolved photoluminescence and RMCD measurements. In both cases, the laser was normally incident on the sample with a beam spot size of about 1 μm diameter.

For photoluminescence measurement, we used 3 µW for spatially resolved maps and 10 µW for magnetic field sweeps. For RMCD measurement, we used the same excitation power as the photoluminescence measurement for spatially resolved maps and magnetic field sweeps. The experimental setup is detailed in Ref 26.

## Acknowledgements

We thank Adrian Lonescu and Ilham Wilson for assistance in sample fabrication. This work was mainly supported by the Department of Energy, Basic Energy Sciences, Materials Sciences and Engineering Division (DE-SC0018171). The understanding of magnetic proximity effect was partially supported by DoE Pro-QM EFRC (DE-SC0019443). Work at HKU is supported by the RGC of HKSAR (17303518P). Work at ORNL (M.A.M.) was supported by the US Department of Energy, Office of Science, Basic Energy Sciences, Materials Sciences and Engineering Division. K.W. and T.T. acknowledge support from the Elemental Strategy Initiative conducted by the MEXT, Japan and the CREST (JPMJCR15F3), JST. K.-M.C.F. and X.L. acknowledge support by University of Washington Innovation Award. X.X. acknowledges the support from the State of Washington funded Clean Energy Institute and from the Boeing Distinguished Professorship in Physics.

## Author Contributions

X.X. W.Y. D.X. conceived the project. D.Z. fabricated the sample. D.Z., K.L.S, L.X. performed the experiment assisted by N.P.W., supervised by X.X. and K.-M.C.F. M.A.M synthesized and characterized the bulk $CrI_3$ crystal. T.T. and K.W. synthesized the bulk h-BN Crystal. D.Z. X.X. W.Y. and D.X. analyzed data. X.X. D.Z. K.L.S. W.Y. and D.X. wrote the paper with input from all authors. All authors discussed the results.

## Competing Interests Statement

The authors declare that they have no competing interests.

**Data Availability:** The data that support the findings of this study are available from the corresponding author upon reasonable request.

**Figures**

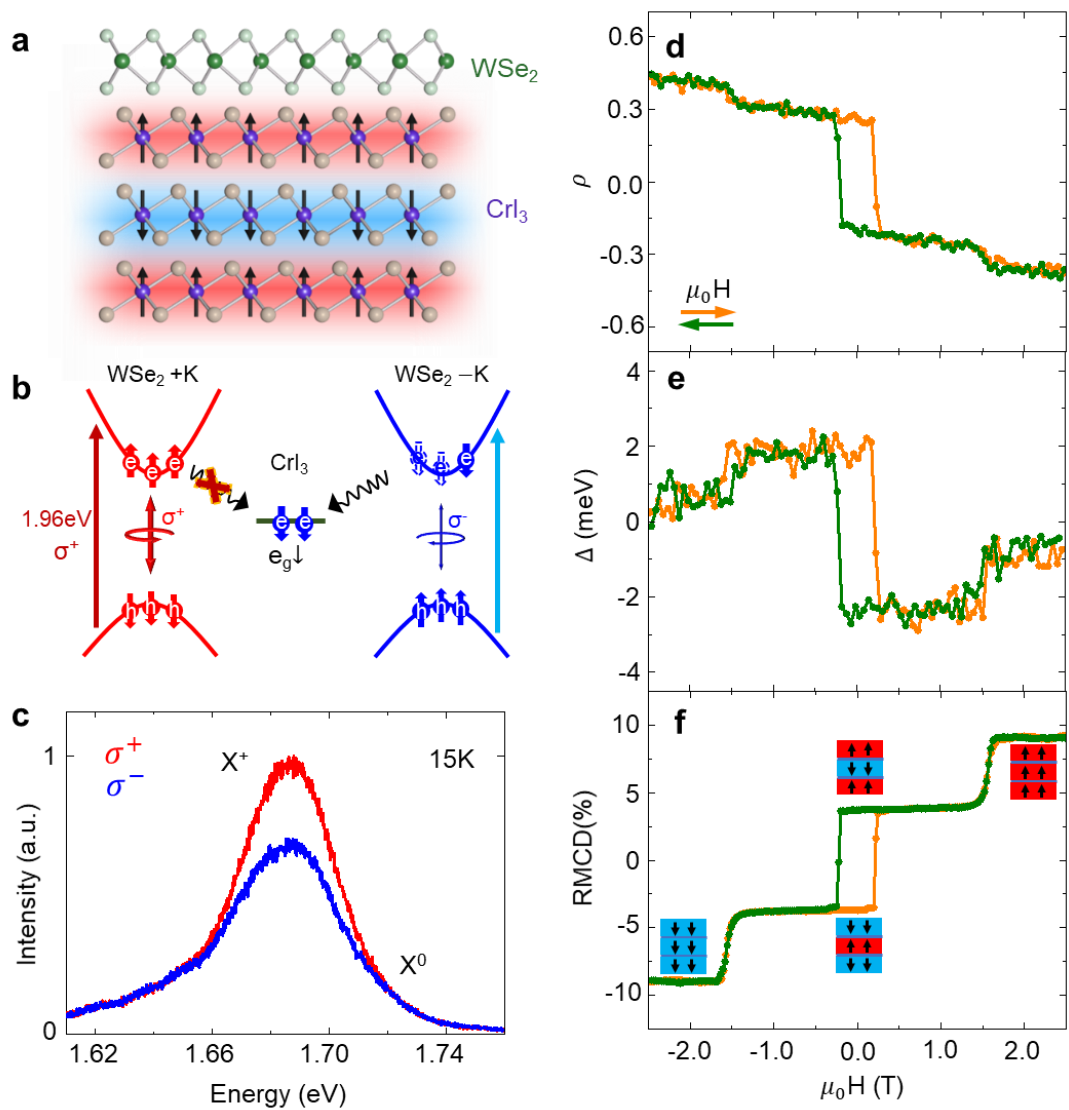

**Figure 1 | Proximity control of valley dynamics in monolayer WSe$_2$ interfacing with trilayer CrI$_3$.**
**a**, Schematic of monolayer WSe$_2$ and trilayer CrI$_3$ heterostructure. **b**, Schematic depicting spin dependent charge transfer. Assuming spin-down magnetization CrI$_3$, the optically excited spin-down electrons in the −K valley will transfer to the Cr$^{3+}$ spin polarized 3d e$_g$ band, while charge transfer of the spin up electron in the +K valley is suppressed. **c**, Polarization-resolved photoluminescence of WSe$_2$/trilayer CrI$_3$ heterostructure at 15 K and zero magnetic field, showing spontaneously $\sigma^+$ polarized photoluminescence. **d**, Degree of circular polarization, **e**, valley Zeeman splitting in photoluminescence and **f**, reflective magneto-circular dichroism (RMCD) as a function of magnetic field. Orange and green curves represents magnetic field sweeping up (increase) and down (decrease), respectively.

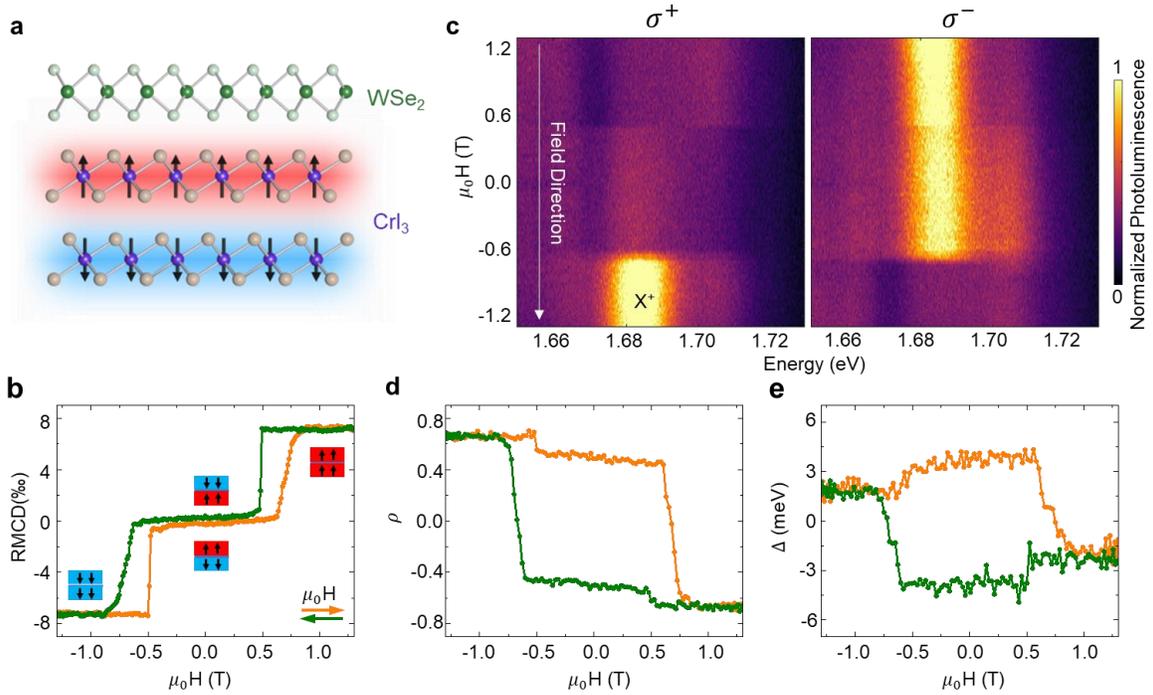

**Figure 2 | Proximity effect in monolayer WSe$_2$/Bilayer CrI$_3$ heterostructure. a**, Schematic of monolayer WSe$_2$ and bilayer CrI$_3$ heterostructure. **b**, RMCD as a function of magnetic field, showing typical features of a layered antiferromagnetic bilayer CrI$_3$. **c**, Photoluminescence intensity plot of co-$\sigma^+$ (left) and co-$\sigma^-$ (right) polarized excitation and detection as a function of magnetic field and photoenergy. **d**, Degree of circular polarization and **e**, valley Zeeman splitting as a function of magnetic field, extracted from data in (c).

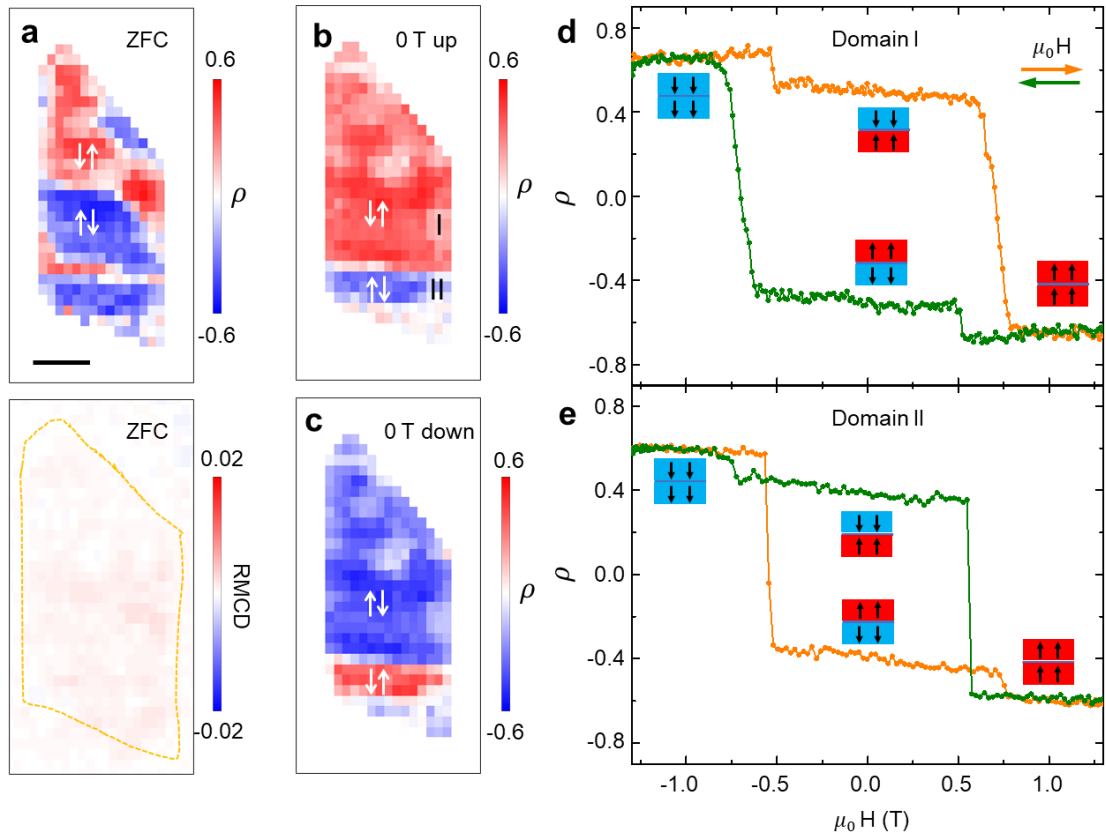

**Figure 3 | Imaging layered antiferromagnetic domains in bilayer CrI$_3$ by monolayer WSe$_2$. a**, Spatially resolved RMCD (bottom) and spontaneous circular polarization $\rho$ of WSe$_2$ photoluminescence (top) as the device BD1 was zero-field-cooled (ZFC) down to 1.6K. Scale bar: 2 μm. **b**, Spatial map of $\rho$ at zero magnetic field as CrI$_3$ is initialized in ↓↓ state by applying negative magnetic field. I and II labels two layered antiferromagnetic domains. **c**, The same as **b** but with CrI$_3$ is initialized at ↑↑ state by applying large positive magnetic field. **d-e**, $\rho$ as a function of magnetic field at selected spots in domain I (d) and II (e). Insets depict the magnetic states.

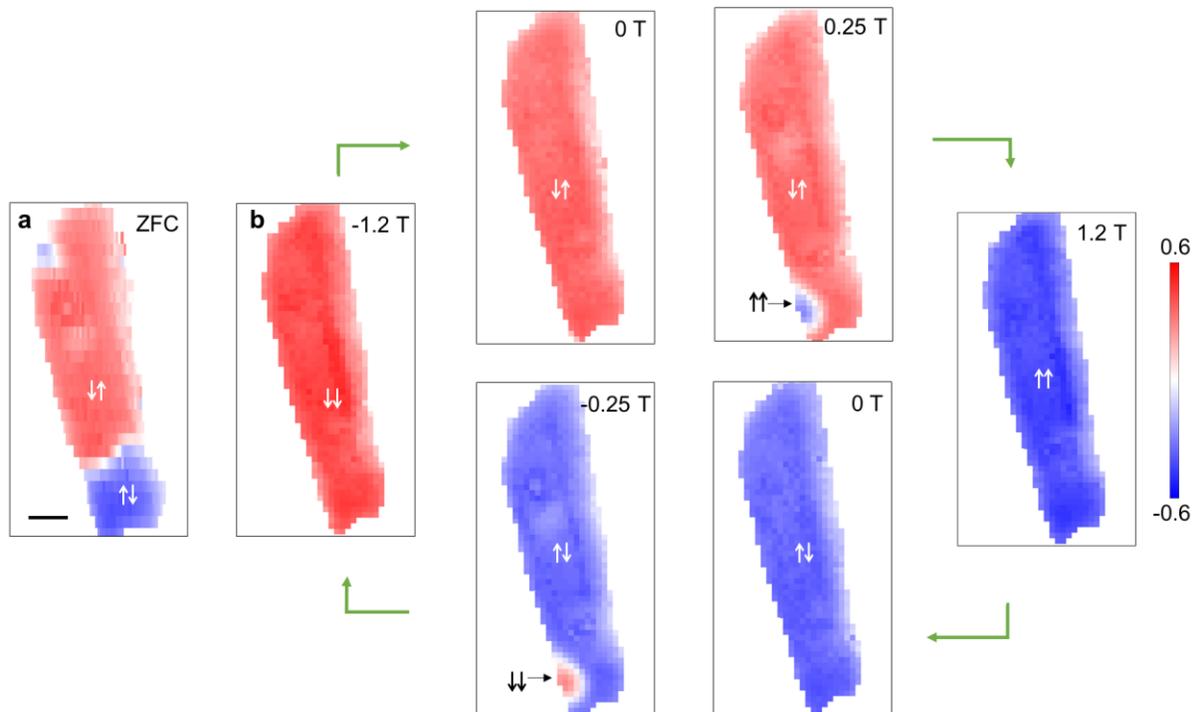

**Figure 4 | Imaging layered antiferromagnetic-ferromagnetic domains in bilayer CrI$_3$ by monolayer WSe$_2$. a**, Spatial map of $\rho$ of spontaneously circularly polarized WSe$_2$ photoluminescence as device BD2 was zero field cooled (ZFC) to 6.6K. Scale bar: 2 μm. **b**, Spatial map of $\rho$ at selected magnetic fields. Layered antiferromagnetic and ferromagnetic domains at ±0.25 T maps are indicated. Insets depict the magnetic states.

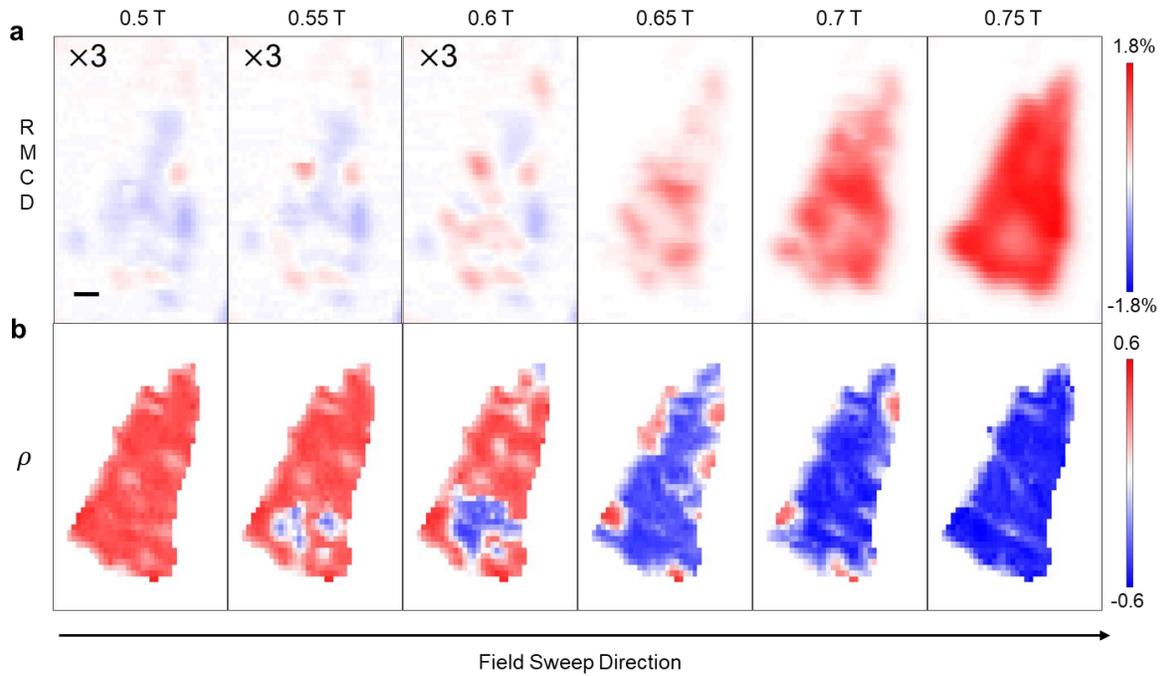

**Figure 5 | Imaging domain dynamics near metamagnetic transitions of bilayer CrI$_3$.** Spatial maps of **a**, RMCD, and **b**, $\rho$ at selected magnetic fields near metamagnetic transition from a third monolayer WSe$_2$/bilayer CrI$_3$ device (BD3) while the magnetic field sweeps up. The RMCD values in the first three maps in **a** are tripled for the better visibility. The evolution of layered antiferromagnetic and fully spin polarized ferromagnetic domains is evident in the maps of $\rho$. Scale bar: 2 μm.